\documentclass[reprint,prl]{revtex4-1}
\usepackage[utf8]{inputenc}
\usepackage{amsmath}
\usepackage{gensymb}
\usepackage{placeins}
\usepackage{graphicx}
\usepackage{natbib}

\begin{document}

\title{Nonreciprocity in Photon Pair Correlations}
\author{Austin Graf$^{\text{ 1,2}}$}
\author{Steven D. Rogers$^{\text{ 3}}$}
\author{Jeremy Staffa$^{\text{ 1,2}}$}
\author{Usman A. Javid$^{\text{ 1,2}}$}
\author{Dana H. Griffith$^{\text{ 4}}$}
\author{Qiang Lin$^{\text{ 1,2,5}}$}
\affiliation{$^1$Institute of Optics, University of Rochester, Rochester, NY 14627}
\affiliation{$^2$Center for Coherence and Quantum Optics, University of Rochester, Rochester, NY 14627}
\affiliation{$^3$John Hopkins University, Applied Physics Laboratory, Laurel, MD 20723}
\affiliation{$^4$Department of Physics, Wellesley College, Wellesley, MA 02841}
\affiliation{$^5$Department of Electrical and Computer Engineering, University of Rochester, Rochester, NY 14627}

\begin{abstract}
\indent Nonreciprocal optical systems have found many applications altering the linear transmission of light as a function of its propagation direction. Here we consider a new class of nonreciprocity which appears in photon pair correlations and not in linear transmission. We experimentally demonstrate and theoretically verify this nonreciprocity in the second-order coherence functions of photon pairs produced by spontaneous four-wave mixing in a silicon microdisk. Reversal of the pump propagation direction can result in substantial extinction of the coherence functions without altering pump transmission.
\end{abstract}

\maketitle

	\indent Nonreciprocal optical systems, which exhibit a change in transmission upon reversing the propagation of an input field, have generated great interest in classical electromagnetism. Many nonreciprocal systems have been achieved with magnetic biasing \cite{EMNonreciprocityReview}, while others have relied on dynamic modulation \cite{NonreciprocalPhotonicsTimeModulation,OptomechanicalNonreciprocityCommentary,OptomechanicalNonreciprocityCoupling}. Optical nonlinearity has also been leveraged to induce nonreciprocity through self-biasing, creating passive optical isolation \cite{OpticalIsolatorNonlinearTopological}. These classical nonreciprocal systems form the foundation of optical isolation, which has proven vital to laser operations \cite{WhatIsAnOpticalIsolator,AFaradayEffectOpticalIsolator1963}. Recently, nonreciprocity has been extended to lasing itself \cite{NonreciprocalLasingInTopologicalCavities}, and even been applied to quantum systems to achieve single-photon routing \cite{PhotonRoutingChiral}. However, the conjunction of nonlinearity and nonreciprocity remains little-explored in the quantum regime, meriting a more detailed investigation. \\
	\indent In classical electromagnetism, the change in the ratio between transmitted and incoming fields that occurs upon swapping sources with detectors determines the extent of a system's nonreciprocity \cite{EMNonreciprocityReview}. Here, we introduce a new class of nonreciprocity, which does not appear in the linear transmission of light, but rather in the quantum measurement of field coherence. Specifically, in a nonlinear system operating in the single-photon regime, nonreciprocity emerges in the second-order coherence function between quantum optical fields. \\
\begin{figure*}[!htpb]
\begin{center}
\includegraphics[width=2\columnwidth]{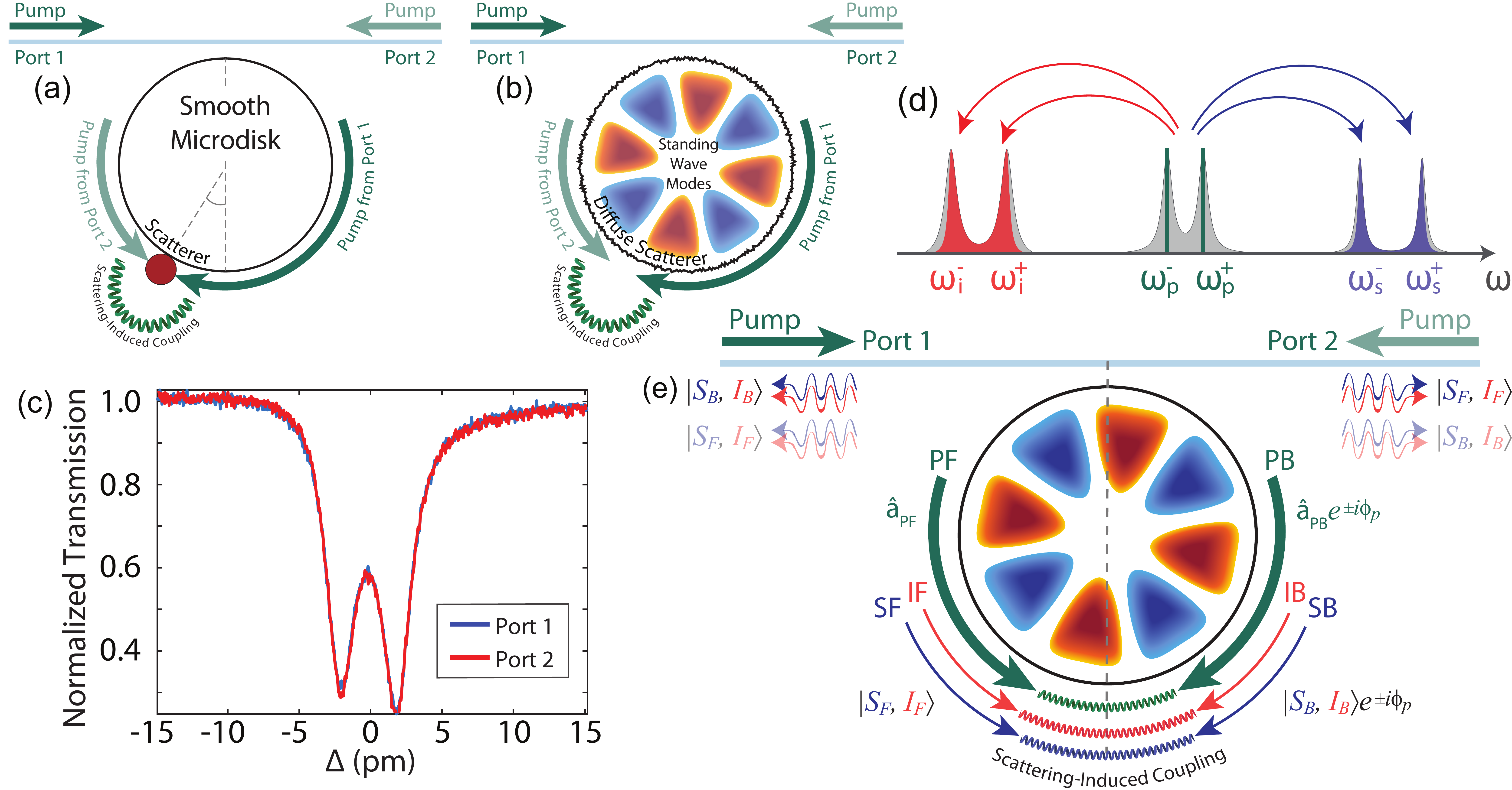}
\caption{Nonreciprocity arises in photon pairs generated by cavity-enhanced spontaneous four-wave mixing (SFWM) via coupled whispering-gallery modes (WGMs). (a) A depiction of a smooth silicon microdisk with a single surface scatterer. Pump laser light (green) is introduced at port 1 and evanescently coupled into the microdisk, exciting a forward-propagating WGM. The scatterer couples this mode to a backward-propagating WGM. If the device were instead pumped from port 2 (translucent green), the backward-propagating WGM would be excited and light could scatter to the forward-propagating WGM. The position of the scatterer crucially affects the relative phase between WGMs in each pumping scenario. (b) A depiction of a rough silicon microdisk. Nanoscale surface roughness acts as a distributed scatterer, coupling WGMs and creating standing wave eigenmodes (blue and orange). (c) The experimental cavity transmission of the pump doublet resonance for a pump field introduced at port 1 (blue) and at port 2 (red), which closely match. (d) A frequency domain representation of SFWM in the silicon microdisk showing several biphoton creation pathways. (e) SFWM in a silicon microdisk with scattering-induced coupling. Generated signal and idler photons may coherently scatter between clockwise- and counterclockwise-propagating modes before exiting the device either forward- or backward-propagating relative to the input pump field. The phase relationship between pump modes and generated biphoton states depends on pump field's initial propagation direction.}
\end{center}
\end{figure*}	
	\indent We begin to explore the aforementioned nonreciprocity by considering first an optical microdisk with a single scatterer on its surface, shown in Fig.~1(a). Laser light is coupled into the cavity in one of two possible propagation directions via port 1 or port 2, exciting a co-propagating whispering-gallery mode (WGM). The single scatterer couples the clockwise-propagating and counter-clockwise-propagating WGMs as light scatters between the modes. The position of the scatterer introduces a crucial asymmetry between ports, creating a dependence of the relative phase between WGMs on the laser light's entry port. \\
	\indent In practice, the disk's surface is not smooth but instead possesses nanoscale roughness, as in Fig.~1(b). The asymmetrical distribution of the surface roughness combines with its sub-wavelength scale to create a system analogous to that of the smooth microdisk with a single scatterer in Fig.~1(a). Scattering-induced coupling again arises between counter-propagating WGMs. With scattering populating both WGMs and breaking their degeneracy, standing wave eigenmodes such as those in Fig.~1(b) form in the microdisk. The asymmetry observed in the case of single scatterer persists in the presence of a distributed scatterer like nanoscale surface roughness. That phase asymmetry is now most easily visualized in the standing wave mode pattern, fixed in place by the surface roughness. \\
	\indent The observed phase asymmetry has no effect on the linear optical response of the system. Fig.~1(c) shows that there is no change in the transmission when the laser light's input direction is changed. However, if the microdisk additionally exhibits resonantly enhanced third-order ($\chi^{(3)}$) nonlinearity, the laser light may act as a pump to generate pairs of signal and idler photons by spontaneous four-wave mixing (SFWM) \cite{BoydNonlinearOptics,SFWMInMicroring,CWPairGenerationSOI}, and nonreciprocity can be observed. Fig.~1(d) shows the nonlinear process in the frequency domain, with each of the standing wave pump modes (green) populated by scattering-induced coupling at eigenfrequencies $\omega_p^{\pm}=\omega_p \pm \lvert \beta \rvert$. Here the central resonance frequency is given by $\omega_p$ while $\lvert \beta \rvert$ denotes the scattering rate, or equivalently half of the doublet resonance splitting \cite{SplittingMieModes1995,SingleRayleighScattererQuantumLight,UnifiedApproachToSplitting}. If nanoscale surface roughness likewise induces splitting at the signal (red) and idler (blue) resonance frequencies, a quantum interference arises between SFWM biphoton creation pathways, an interference that critically depends upon the pump field's incoming propagation direction. \\
	\indent The system in Fig.~1(e) shows how modal coupling at all three (pump, signal, and idler) resonances affects the nonlinear generation of signal and idler photon pairs. Modal coupling dramatically alters the photons' dynamics. Pump light undergoing resonantly-enhanced SFWM will always produce co-propogating photon pairs to conserve momentum. However, in the presence of modal coupling, these photons can coherently scatter between clockwise- and counterclockwise-propagating WGMs. With respect to the WGMs, this coherent scattering creates a time-evolving path entanglement within the cavity, as the propagation direction of either photon in an entangled pair oscillates via scattering. Oscillations in propagation direction persist until each photon exits the cavity in either the same direction as the input pump (forward) or the opposite direction (backward). \\
\begin{figure*}[!htpb]
\begin{center}
\includegraphics[width=2\columnwidth]{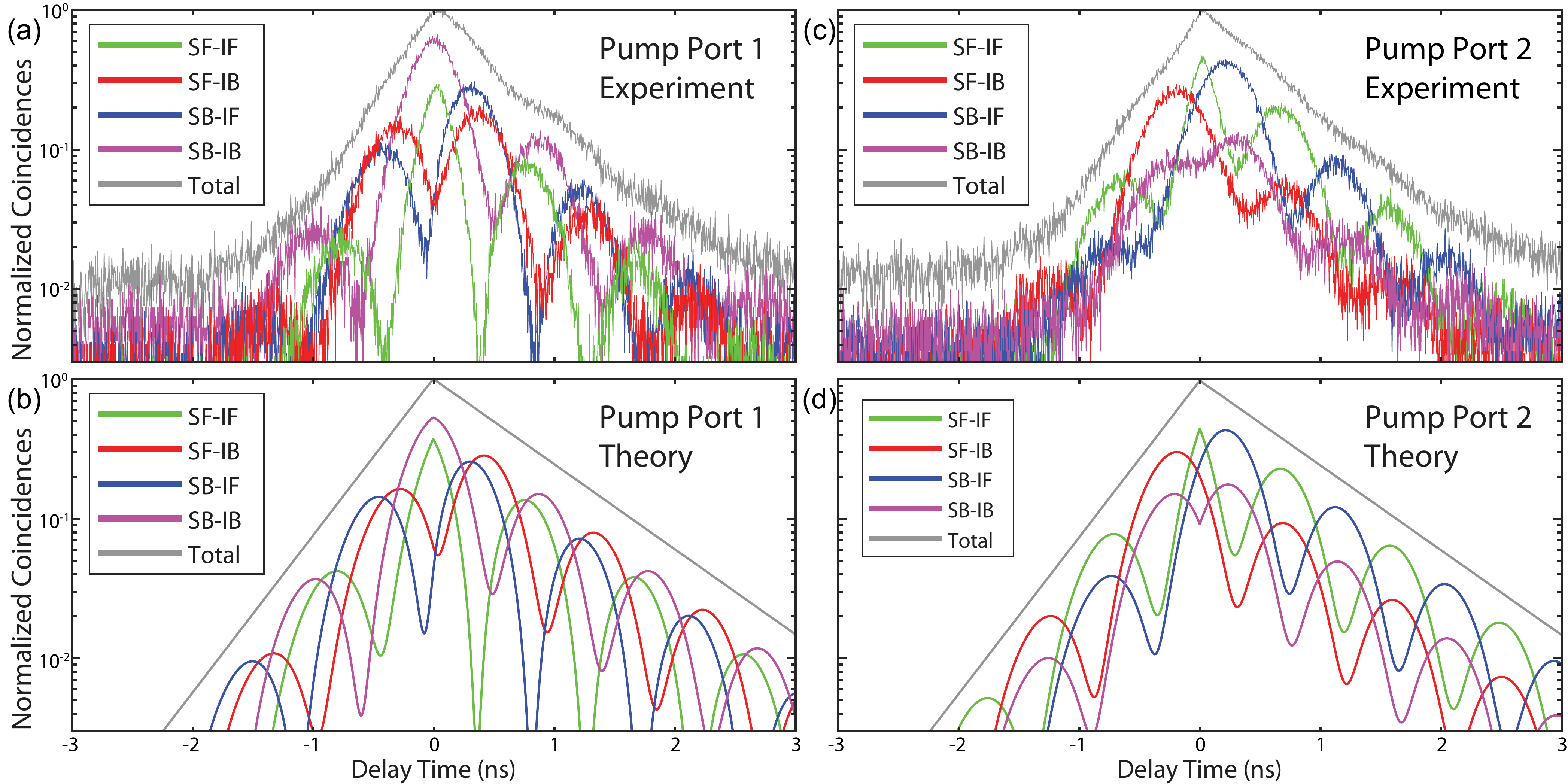}
\caption{Normalized second-order biphoton coherence functions. (a) and (b) Measured coincidences are plotted as a function of the difference in the detection times of the signal and idler photons produced by a pump input at port 1 (a) and port 2 (b) for all possible biphoton path configurations: signal-forward and idler-forward (SF-IF), signal-forward and idler-backward (SF-IB), signal-backward and idler-forward (SB-IF), and signal-backward and idler-backward (SB-IB). The sum of coherence functions for all four path configurations gives the total coherence function, and all coherence functions are normalized to the maximum of the respective total coherence function. (d) and (e) The theoretically obtained coherence functions corresponding to pump port 1 and pump port 2, respectively.}
\end{center}
\end{figure*}
	\indent Each signal and idler photon exits the optical cavity probabilistically, at a rate governed by the cavity quality factor. If the signal exits the cavity first, then the idler is free to coherently scatter within the optical cavity. This coherent scattering alters the idler propagation direction correlated with the signal photon's exit direction. An illustrative example considers the situation in which the signal and idler are created propagating in the clockwise direction, and the signal exits the cavity in the forward direction while the idler remains in the cavity. The continued scattering of the idler in the cavity will cause this forward-propagating signal photon to oscillate between correlation with a clockwise-propagating idler and correlation with a counterclockwise-propagating idler until the idler leaves the cavity. The cross-correlation of the photon pair is a function of the difference in emission time between the signal and idler photons, $\tau \equiv t_s-t_i$. As such, oscillations appear in this biphoton coherence function, with an oscillation rate equal to the scattering rate $ \lvert \beta \rvert$. These oscillations can equivalently be described in the frequency domain as a beating between signal and idler eigenfrequencies, with the beat frequency given by half the mode splitting, again giving  $ \lvert \beta \rvert$. \\
	\indent The oscillations are evident in the biphoton coherence functions for four possible pairs of propagation directions: signal-forward and idler-forward (SF-IF), signal-forward and idler-backward (SF-IB), signal-backward and idler-forward (SB-IF), and signal-backward and idler-backward (SB-IB). Summing these four functions yields the exponential decay envelope one would expect from a cavity in which no modal coupling is present, verifying conservation of total probability. For a pair of signal and idler photon paths, denoted $j = f,b$ and $k = f,b$ respectively, the probability that the signal photon arrives at time $t_{sj}$ and the idler photon at time $t_{ik}$ is given by \cite{CommPhysLinGroupPaper}
\begin{align}
p(t_{sj},&t_{ik})= \nonumber \\
&N e^{-\Gamma_{tm} \lvert \tau_{jk} \rvert} \lvert \zeta_m^{jk} \cos (\beta_m \tau_{jk})+\eta_m^{jk} sin  (\beta_m \tau_{jk})\rvert^2
\end{align}
The subscript $m=s$ for $t_{sj}>t_{ik}$ and $i$ for $t_{sj}<t_{ik}$. The total decay rate $\Gamma_{tm}$ is consequently determined from either the signal or idler decay rate, depending on the value of $m$. The complex scattering rate $\beta_m$ likewise differs between the signal and idler, and it may be written $\beta_m = \lvert \beta_m \rvert e^{i \phi_m}$ where $\phi_m$ is the phase accrued upon scattering. The normalization coefficient $N$ is determined by parameters of the device. Meanwhile, $\zeta_m^{jk}$ and $\eta_m^{jk}$ are functions of the energy and relative phase of the clockwise- and counterclockwise-propagating intracavity pump modes shown in Fig.~2.\\
\begin{figure}[!htbp]
\begin{center}
\includegraphics[width=\columnwidth]{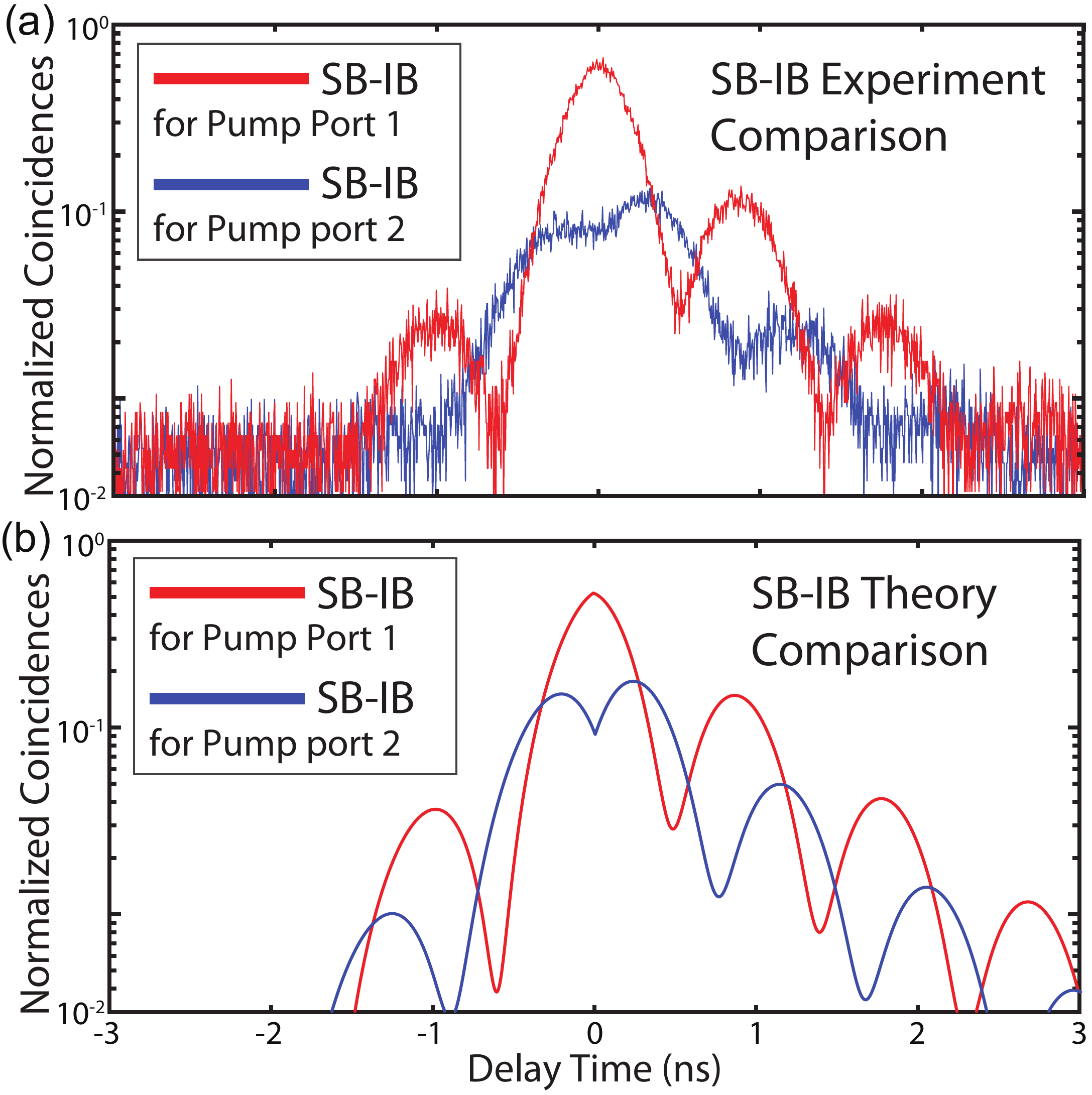}
\caption{Normalized second-order biphoton coherence functions. (a) A direct comparison of signal-backward and idler-backward (SB-IB) normalized second-order biphoton coherence functions produced by a pump input at port 1 (red) and port 2 (blue). (b) Theoretically obtained coherence functions analogous to those in (a).}
\end{center}
\end{figure}
	\indent The complex amplitude of these intracavity pump fields can be found as a function of the laser detuning from resonance $\Delta$ by solving the coupled-mode equations in the steady state. For an input field $b_1$ introduced at port 1, the forward and backward propagating fields, relative to the pump direction, are 
\begin{align}
a_{f}(\Delta) &= \kappa(\Delta) (i\Delta-\Gamma_{tp}/2)b_1\\
a_{b}(\Delta) &= \kappa(\Delta)(-i\lvert \beta_p \rvert e^{-i \phi_p})b_1
\end{align} 
where $\kappa (\Delta)$ depends upon the cavity decay rate and the modal scattering rate. Pumping at port 2 with field $b_2$, the complex mode amplitudes are instead 
\begin{align}
a_{f}(\Delta) &= \kappa(\Delta) (i\Delta-\Gamma_{tp}/2)b_2\\
a_{b}(\Delta) &= \kappa(\Delta)(-i\lvert \beta_p \rvert e^{i \phi_p})b_2
\end{align}
Thus pumping at the second port instead of the first alters the relative phase between intracavity modes by $2\phi_p$, twice the pump scattering phase. The phase difference will influence the quantum interference occurring within the microresonator and induce corresponding phase shifts in the second-order biphoton cross-correlations.\\
	\indent A silicon microdisk of radius 4.5 $\mu$m and thickness 260 nm is used to experimentally verify that non-reciprocity arises in the second-order cross-correlation. The SFWM process harnesses three adjacent quasi-transverse-magnetic (quasi-TM) doublet modes at wavelengths 1532, 1551, and 1569 nm, and average intrinsic quality factors over $8\times 10^5$. The magnitude of the complex scattering rate is 0.55, 0.26, and 0.48 GHz for signal, pump, and idler modes respectively.\\
	\indent Pump light is evanescently coupled into the device via either port 1 or port 2 of the tapered optical fiber. Wavelength-division multiplexers separate signal, idler, and pump photons. Optical switches allow for control over which propagation pathway of emitted signal and idler photons are detected with superconducting nanowire single-photon detectors (SNSPDs). In this way all combinations of propagation directions for signal and idler photon pairs can be measured. A time-correlated single-photon counter (TCSPC) is used to acquire the detection times of the signal and idler photons. \\
	\indent Fig.~2(a) and Fig.~2(b) contrasts the biphoton correlation functions that arise in the presence of a pump laser from port 1 with those generated by a pump laser from port 2. Normalized biphoton coincidence counts collected over a three minute data acquisition time are plotted against delay time $\tau$. A dramatic phase shift can be observed in the correlations' oscillations upon reversing the pump direction. Theory derived from the coupled-mode equations and Heisenberg-Langevin equations \cite{HeisenbergLangevinVSQuantumMasterEquation} predicts this phase shift.  Fig.~2(c) and Fig.~2(d) demonstrate the correspondence with theory, respectively for inputs at pump port 1 and pump port 2. Fig.~3(a) directly compares the SB-IB path configuration's biphoton coherence functions for each pump port case while Fig.~3(b) shows the theoretical results for these configurations. In all cases, the theory accurately replicates the experimentally observed system behavior. \\
\begin{figure}[!htbp]
\begin{center}
\includegraphics[width=\columnwidth]{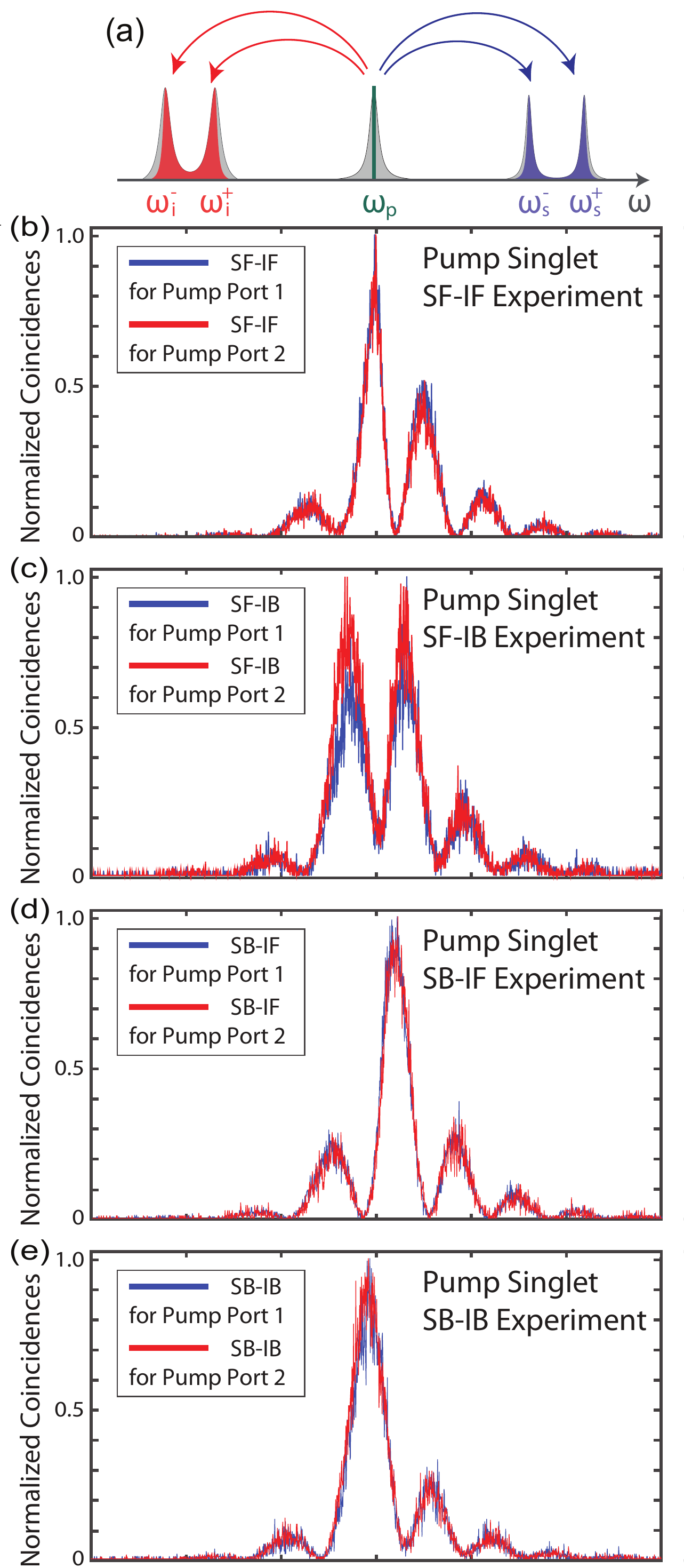}
\caption{Description of a microdisk with a singlet pump mode. (a) A frequency domain picture of SFWM with a singlet pump. (b) Second-order coherence function (normalized to its maximum) for (b) SF-IF, (c) SF-IB, (d) SB-IF, and (e) SB-IB biphoton path configurations from a microdisk with a singlet pump mode.}
\end{center}
\end{figure}
	\indent Further evidence verifying the mechanism of the biphoton coherence functions' phase shifts can be provided by examining a second device, one which does not exhibit coupling between pump modes. Such a device, the resonance structure of which is presented in Fig.~4(a), would still show oscillations in the biphoton coherence function as a result of modal coupling in the signal and idler modes, but these oscillations would not change with a change in pump input direction. This implies that the doublet nature of the pump mode is indeed vital to the observed phase shift. \\
	\indent We tested such a device, and the results are displayed in Fig.~4(b-e). For all four biphoton path configurations, there is no change observed in the second-order coherence function upon changing pump direction. This verifies that the coupling between counter-propagating pump modes is requisite to the nonreciprocity observed in the second-order biphoton coherence functions.\\
	\indent The silicion microdisk's intrinsic scattering phase primarily determines the extent of the phase shift that occurs in the biphoton coherence functions when the input pump port is changed. In the described experiment, the coupling mechanism between counter-propagating modes is Rayleigh scattering mediated by the device's nanoscale surface roughness, and the associated scattering phase is not tuned. Importantly though, this phase can be tuned for the desired application by modulating the device's radius during fabrication \cite{LinGroupResonanceEngineering}, or, if reconfigurability is required, by applying an external scattering probe \cite{ControlledModeSplitting,NearFieldMappingMicroresonators}. \\
	\indent The degree of the phase shift, and likewise the extinction ratio, can also be tuned by varying the pump laser wavelength. Detuning the pump laser from resonance changes the relative power and relative phase between counter-propagating pump fields in the device, which can shift peaks to nulls (and nulls to peaks) in the second-order biphoton coherence functions. \\
	\indent The extent of the device's tunability enhances its usefulness. Under post-selection, it can provide a highly configurable source of path-entangled photon pairs. The introduction of active means of device tuning such as phase modulation of the pump laser or mechanical modulation of the device could lessen the need for post-selection by providing greater control over the biphoton quantum state, resulting in a source that could provide a unique path entanglement depending on the pump propagation direction. \\
	\indent Beyond its applicability, the silicon microdisk system illustrates a new class of nonreciprocity in the second-order coherence functions of its nonlinearly created photon pairs. The complete experimental and theoretical description of such a system provides deeper insight and intuition into the quantum behavior of nonlinear nonreciprocal systems, inviting the possibility of new developments. Silicon photonic structures themselves provide a compelling platform with which to investigate and advance nonlinear nonreciprocal optics. Moreover, the understanding developed by the theoretical and experimental demonstration of nonreciprocity in biphoton coherence functions can be applied to any system regardless of platform.
\begin{acknowledgments}
This work is supported in part by National Science Foundation (NSF) under grant No. EFMA-1641099, ECCS-1810169, and ECCS-1842691. The work was performed in part at the Cornell NanoScale Facility, a member of the National Nanotechnology Coordinated Infrastructure (NNCI), which is supported by the National Science Foundation (Grant NNCI-2025233).
\end{acknowledgments}

\section{Apendices}

\subsection{A. Theoretical Description of The Microdisk System}
\indent The theoretical description of the microdisk system is adapted from that of our previous work \cite{CommPhysLinGroupPaper}. An error which appeared in Equations 27-29 of the supplementary material, altering Equations 40-43 of the supplementary material, has been herein corrected.\\
\indent The Hamiltonian $H = H_0+H_I$ given in Equations 6-7 describes the creation and evolution of photon pairs via spontaneous four-wave mixing (SFWM) in the presence of modal coupling in an optical microdisk cavity. The resonantly enhanced nonlinear interaction encompasses pump (p), signal (s), and idler (i) whispering gallery modes (WGMs). For m$=$p,s,i, the pump, signal, and idler modes respectively have resonance frequencies $\omega_{0\text{m}}$, external coupling rate $\Gamma_\text{em}$, and modal splittings $2\lvert\beta_\text{m}\rvert$ where $\beta_\text{m}\equiv \lvert\beta_\text{m}\rvert e^{i\phi_\text{m}}$ is the complex Rayleigh scattering rate. Creation and annihilation operators $b^\dagger_\text{mf}$ and $b_\text{mf}$ refer to forward propagation in a tapered optical waveguide evanescently coupled to the optical microdisk cavity. For the purpose of this theoretical description, all directions forward and backward are defined with respect to port 1. Analogous operators $b^\dagger_\text{mb}$ and $b_\text{mb}$ refer to backward propagation. Likewise creation and annihilation operators within the microdisk are denoted $a^\dagger_\text{mf}$, $a^\dagger_\text{mb}$, $a_\text{mf}$, $a_\text{mb}$. Intracavity creation and annihilation operators are normalized such that $a^\dagger_\text{mj}a_\text{mj}=N_\text{mj}$, the photon number operator for mode m$=$p,s,i and propagation direction j$=$f,b. The external operators $b_\text{mj}$ are instead normalized such that $b^\dagger_\text{mj}b_\text{mj}$ denotes the photon flux into the cavity. Additionally, $\left[b_\text{mj}(t),b^\dagger_\text{m'j'}(t')\right]=\delta_\text{mm'}\delta_\text{jj'}\delta\left(t-t'\right)$. The vacuum coupling rate for self-phase modulation (SPM) is $g_\text{p}$ while that for cross-phases modulation (XPM) is $g_\text{pm}$ with m$=$s,i denoting either the signal or idler mode. The vacuum coupling rate for SFWM is $g_\text{psi}$. \\
\indent The first four terms of $H_0$ (first line of Equation 6) provide the linear Hamiltonian of the closed system, that is a microdisk with no input or output channels or nonlinearity. The third and fourth terms in particular indicate the presence of modal splitting, which shifts the system's eigenfrequencies by $\pm\lvert\beta_\text{m}\rvert$ and creates standing-wave eigenmodes. The final four terms of $H_0$ (second line of Equation 6) introduce a channel by which photons can enter and leave the microdisk, opening the system. \\
\indent The interaction Hamiltonian $H_I$ encompasses all relevant Kerr nonlinearity present in the system. The first three terms of $H_I$ (first line of Equation 7) provide for SPM of the pump fields. The second line of Equation 7 likewise provides for XPM between the pump field and the signal and idler fields. Because the pump field contains far more photons than the signal and idler fields, SPM and XPM within and between signal and idler fields have been ignored. The final line of Equation 7 contains all terms associated with SFWM. \\
\begin{widetext}
\begin{gather}
\begin{align}
H_0 = & \sum_\text{m=p,s,i} \hbar \omega_\text{0m} \left(a^\dagger_\text{mf} a_\text{mf} + a^\dagger_\text{mb}  a_\text{mb} \right) - \left( \hbar \beta_\text{m} a^\dagger_\text{mf} a_\text{mb} + \hbar \beta^*_\text{m} a^\dagger_\text{mb} a_\text{mb} \right) \nonumber \\
& -\hbar \sqrt{\Gamma_\text{em}} \left[ \left(a^\dagger_\text{mf}b_\text{mf} + a^\dagger_\text{mb}b_\text{mb} \right) e^{-i\omega_\text{m}} +  \left(b^\dagger_\text{mf}a_\text{mf} + b^\dagger_\text{mb}a_\text{mb} \right) e^{i\omega_\text{m}}  \right]  \\
H_I = & \frac{\hbar g_\text{p}}{2} \left[ \left( a^\dagger_\text{pf}\right)^2 a^2_\text{pf}+\left(a^\dagger_\text{pb}\right)^2 a^2_\text{pb} + 4a^\dagger_\text{pf}a_\text{pf}a^\dagger_\text{pb}a_\text{pb}\right] \nonumber \\
& +2\hbar \left(a^\dagger_\text{pf}a_\text{pf} + a^\dagger_\text{pb} a_\text{pb}\right) \left[ g_\text{ps} \left(a^\dagger_\text{sf} a_\text{sf} + a^\dagger_\text{sb} a_\text{sb} \right) + g_\text{pi} \left(a^\dagger_\text{if} a_\text{if} + a^\dagger_\text{ib} a_\text{ib} \right) \right] \nonumber \\
& + \hbar g_\text{psi} \left[ a^\dagger_\text{sf} a^\dagger_\text{if} a^2_\text{pf} + a^\dagger_\text{sb} a^\dagger_\text{ib} a^2_\text{pb} \right] + \hbar g^*_\text{psi} \left[ \left( a^\dagger_\text{pf}\right)^2 a_\text{sf} a_\text{if} + \left( a^\dagger_\text{pb}\right)^2 a_\text{sb} a_\text{ib} \right]
\end{align}
\end{gather}
\end{widetext}
\indent Expressions for the biphoton coherence functions of the signal and idler photon pairs generated by SFWM can be obtained by solving the Heisenberg-Langevin equations of motion). Because the frequencies and field profiles of the pump, signal, and idler are very similar, we here introduce the approximation $g_\text{p} \approx g_\text{pm} \approx g_\text{psi} \equiv g$ where $g = \frac{\hbar c \eta n_2 \omega_\text{p} \sqrt{\omega_\text{s} \omega_\text{i}}}{n_s n_i V}$. Here, $\eta$ is the fractional spatial overlap between modes, $n_2 = \frac{3\chi^{(3)}}{4\epsilon_0cn^2_\text{p}}$ is the Kerr nonlinear index \cite{BoydNonlinearOptics}, $n_\text{s}$ and $n_\text{i}$ are the refractive indices at the signal and idler wavelengths, respectively, and $V$ is the effective mode volume. The Heisenberg-Langevin equations are then
\begin{widetext}
\begin{gather}
\begin{align}
&\frac{da_\text{pf}}{dt}=\left(-i\omega_{0\text{p}}-\Gamma_\text{tp}/2\right)a_\text{pf}+i\beta_\text{p}a_\text{pb}-ig\left(a^\dagger_\text{pf}a_\text{pf}+2a^\dagger_\text{pb}a_\text{pb}\right)a_\text{pf}+\zeta_\text{pf}(t), \\
&\frac{da_\text{pb}}{dt}=\left(-i\omega_{0\text{p}}-\Gamma_\text{tp}/2\right)a_\text{pb}+i\beta^*_\text{p}a_\text{pf}-ig\left(a^\dagger_\text{pb}a_\text{pb}+2a^\dagger_\text{pf}a_\text{pf}\right)a_\text{pb}+\zeta_\text{pb}(t), \\
&\frac{da_\text{sf}}{dt}=\left(-i\omega_{0\text{s}}-\Gamma_\text{ts}/2\right)a_\text{sf}+i\beta_\text{s}a_\text{sb}-2ig\left(a^\dagger_\text{pf}a_\text{pf}+a^\dagger_\text{pb}a_\text{pb}\right)a_\text{sf}-iga^\dagger_\text{if}a^2_\text{pf}+\zeta_\text{sf}(t), \\
&\frac{da_\text{sb}}{dt}=\left(-i\omega_{0\text{s}}-\Gamma_\text{ts}/2\right)a_\text{sb}+i\beta^*_\text{s}a_\text{sf}-2ig\left(a^\dagger_\text{pf}a_\text{pf}+a^\dagger_\text{pb}a_\text{pb}\right)a_\text{sb}-iga^\dagger_\text{ib}a^2_\text{pb}+\zeta_\text{sb}(t), \\
&\frac{da_\text{if}}{dt}=\left(-i\omega_{0\text{i}}-\Gamma_\text{ti}/2\right)a_\text{if}+i\beta_\text{i}a_\text{ib}-2ig\left(a^\dagger_\text{pf}a_\text{pf}+a^\dagger_\text{pb}a_\text{pb}\right)a_\text{if}-iga^\dagger_\text{sf}a^2_\text{pf}+\zeta_\text{if}(t), \\
&\frac{da_\text{ib}}{dt}=\left(-i\omega_{0\text{i}}-\Gamma_\text{ti}/2\right)a_\text{ib}+i\beta^*_\text{i}a_\text{if}-2ig\left(a^\dagger_\text{pf}a_\text{pf}+a^\dagger_\text{pb}a_\text{pb}\right)a_\text{ib}-iga^\dagger_\text{sb}a^2_\text{pb}+\zeta_\text{ib}(t),
\end{align}
\end{gather}
\end{widetext}
where the total decay rate $\Gamma_\text{tm}\equiv \Gamma_\text{em} + \Gamma_\text{0m}$ and the operator $\zeta_\text{mj}(t)\equiv i \sqrt{\Gamma_\text{em}} b_\text{mj} + \sqrt{\Gamma_\text{0m}}u_\text{mj}(t)$. The noise operator $u_\text{mj}$ arise from the intrinsic loss of the system. Their commutation relation is analogous to that of the external field operators $b_\text{mj}$:
$\left[u_\text{mj}(t),u^\dagger_\text{m'j'}(t')\right]=\delta_\text{mm'}\delta_\text{jj'}\delta\left(t-t'\right)$.\\
\indent We solve the Heisenberg-Langevin equations under the assumption of a classical, undepleted pump. To begin, we transform the equations into a rotating frame at the pump carrier frequency. This allows us to eliminate SPM and XPM terms, giving
\begin{widetext}
\begin{gather}
\begin{align}
&\frac{da_\text{pf}}{dt}=\left(i\Delta-\Gamma_\text{tp}/2\right)a_\text{pf}+i\beta_\text{p}a_\text{pb}+i\sqrt{\Gamma_\text{ep}}b_\text{pf}, \\
&\frac{da_\text{pb}}{dt}=\left(i\Delta-\Gamma_\text{tp}/2\right)a_\text{pb}+i\beta^*_\text{p}a_\text{pf}+i\sqrt{\Gamma_\text{ep}}b_\text{pb}, \\
&\frac{da_\text{sf}}{dt}=\left(i\Delta-\Gamma_\text{ts}/2\right)a_\text{sf}+i\beta_\text{s}a_\text{sb}-iga^\dagger_\text{if}a^2_\text{pf}+\zeta_\text{sf}(t), \\
&\frac{da_\text{sb}}{dt}=\left(i\Delta-\Gamma_\text{ts}/2\right)a_\text{sb}+i\beta^*_\text{s}a_\text{sf}-iga^\dagger_\text{ib}a^2_\text{pb}+\zeta_\text{sb}(t), \\
&\frac{da_\text{if}}{dt}=\left(i\Delta-\Gamma_\text{ti}/2\right)a_\text{if}+i\beta_\text{i}a_\text{ib}-iga^\dagger_\text{sf}a^2_\text{pf}+\zeta_\text{if}(t), \\
&\frac{da_\text{ib}}{dt}=\left(i\Delta-\Gamma_\text{ti}/2\right)a_\text{ib}+i\beta^*_\text{i}a_\text{if}-iga^\dagger_\text{sb}a^2_\text{pb}+\zeta_\text{ib}(t).
\end{align}
\end{gather}
\end{widetext}
Here, $\Delta \equiv \omega_\text{p} - \omega_\text{0p}$ is the laser frequency detuning from the pump resonance frequency. We can immediately solve for the pump fields from Equations 14-15. In the steady state, we have
\begin{align}
&0=\left(i\Delta -\Gamma_\text{tp}/2 \right) a_\text{pf} +i\beta_\text{p}a_\text{pb}+i\sqrt{\Gamma_\text{ep}}b_\text{pf},\\
&0=\left(i\Delta -\Gamma_\text{tp}/2 \right) a_\text{pb} +i\beta^*_\text{p}a_\text{pf}+i\sqrt{\Gamma_\text{ep}}b_\text{pb}.
\end{align}
Solving for the operators $a_\text{pf}$ and $a_\text{pb}$ as functions of $\Delta$ gives 
\begin{widetext}
\begin{gather}
\begin{align}
a_\text{pf}(\Delta) &= \frac{-i\sqrt{\Gamma_\text{ep}}}{\left(i\Delta-\Gamma_\text{tp}/2\right)^2+\lvert\beta_\text{p}\rvert^2} \left[\left(i\Delta-\Gamma_\text{tp}/2\right)b_\text{pf}-i\beta_\text{p}b_\text{pb}\right],\\
a_\text{pb}(\Delta) &= \frac{-i\sqrt{\Gamma_\text{ep}}}{\left(i\Delta-\Gamma_\text{tp}/2\right)^2+\lvert\beta_\text{p}\rvert^2} \left[\left(i\Delta-\Gamma_\text{tp}/2\right)b_\text{pb}-i\beta^*_\text{p}b_\text{pf}\right].
\end{align}
\end{gather}
\end{widetext}
The final terms of Equations 22-23 differ by a phase factor of $e^{2i\phi_\text{p}}$ due to the complex conjugation of beta that appears in Equation 23. \\
\indent To solve Equations 16-19, we Fourier transform to the frequency domain. Equations 18-19 are complex conjugated such that only intracavity annihilation operators appear for the signal $(a_\text{sf}, a_\text{sb})$ and only intracavity creation operators appear for the idler $(a^\dagger_\text{if}, a^\dagger_\text{ib})$. This gives
\begin{widetext}
\begin{gather}
\begin{align}
&\tilde{\zeta}_\text{sf}(\omega)=\left[-i\left(\omega+\Delta\right)+\Gamma_\text{ts}/2\right]\tilde{a}_\text{sf}(\omega)-i\beta_\text{s}\tilde{a}_\text{sb}(\omega)+iga^2_\text{pf}\tilde{a}^\dagger_\text{if}(-\omega), \\
&\tilde{\zeta}_\text{sb}(\omega)=\left[-i\left(\omega+\Delta\right)+\Gamma_\text{ts}/2\right]\tilde{a}_\text{sb}(\omega)-i\beta^*_\text{s}\tilde{a}_\text{sf}(\omega)+iga^2_\text{pb}\tilde{a}^\dagger_\text{ib}(-\omega), \\
&\tilde{\zeta}^\dagger_\text{if}(-\omega)=\left[i\left(\omega+\Delta\right)+\Gamma_\text{ti}/2\right]\tilde{a}^\dagger_\text{if}(-\omega)+i\beta^*_\text{i}\tilde{a}^\dagger_\text{ib}(-\omega)-ig^*(a^*_\text{pf})^2\tilde{a}_\text{sf}(\omega), \\
&\tilde{\zeta}^\dagger_\text{ib}(-\omega)=\left[i\left(\omega+\Delta\right)+\Gamma_\text{ti}/2\right]\tilde{a}^\dagger_\text{ib}(-\omega)+i\beta_\text{i}\tilde{a}^\dagger_\text{if}(-\omega)-ig^*(a^*_\text{pb})^2\tilde{a}_\text{sb}(\omega).
\end{align}
\end{gather}
\end{widetext}
We rewrite this system as a matrix equation \\$\vec{\zeta}\left(\omega\right)= \textbf{M}\vec{a}\left(\omega\right)$, and expanding, this is
\newpage
\begin{widetext}
\begin{gather}
\begin{pmatrix}
\tilde{\zeta}_\text{sf}(\omega)\\
\tilde{\zeta}_\text{sb}(\omega)\\
\tilde{\zeta}^\dagger_\text{if}(-\omega)\\
\tilde{\zeta}^\dagger_\text{ib}(-\omega)
\end{pmatrix} =
\begin{pmatrix}
-i\left(\omega+\Delta\right)+\Gamma_\text{ts}/2 & -i\lvert\beta_\text{s}\rvert e^{i\phi_\text{s}} & iga^2_\text{pf} & 0 \\
-i\lvert\beta_\text{s}\rvert e^{-i\phi_\text{s}} & -i\left(\omega+\Delta\right)+\Gamma_\text{ts}/2 & 0 & iga^2_\text{pb} \\
-ig^*(a^*_\text{pf})^2 & 0 & i\left(\omega+\Delta\right)+\Gamma_\text{ti}/2 & i \lvert\beta_\text{i} \rvert e^{-i\phi_\text{i}}\\
0 & -ig^*(a^*_\text{pb})^2 & i \lvert \beta_\text{i} \rvert e^{i\phi_\text{i}} & i\left(\omega+\Delta\right)+\Gamma_\text{ti}/2
\end{pmatrix}
\begin{pmatrix}
\tilde{a}_\text{sf}(\omega)\\
\tilde{a}_\text{sb}(\omega)\\
\tilde{a}^\dagger_\text{if}(-\omega)\\
\tilde{a}^\dagger_\text{ib}(-\omega)
\end{pmatrix}.
\end{gather}
\end{widetext}
Here we have written $\beta_\text{m}$ as $\lvert \beta_\text{m} \rvert e^{i\phi_\text{m}}$ to show the phase dependence of the system explicitly. The intracavity field operators can be determined by inverting the matrix $\textbf{M}$: $\vec{a}\left(\omega\right)=\textbf{M}^{-1}\vec{\zeta}\left(\omega\right)$. \\
\indent To determine the second-order cross-correlation functions of the signal and idler photon pairs emitted from the system, we must first introduce the transmitted field operators, $c_\text{mj}$, which are given by the input-output relation $c_\text{mj} = b_\text{mj} +i\sqrt{\Gamma_\text{em}} a_\text{mj}$ \cite{WallsMilburn}. Then the correlation functions for signal photon propagating in direction j=f,b and an idler photon propagating in direction k=f,b is
\begin{widetext}
\begin{gather}
\begin{align}
\nonumber p_\text{jk}\left(t_\text{sj},t_\text{ik}\right) &\equiv \langle c^\dagger_\text{ik}(t)c^\dagger_\text{sj}(t+\tau_\text{jk})c_{sj}(t+\tau_\text{jk})c_\text{ik}(t)\rangle-\langle c^\dagger_\text{sj}(t)c_\text{sj}(t)\rangle \langle c^\dagger_\text{ik}(t)c_\text{ik}(t)\rangle \\
&= \Gamma_\text{es} \Gamma_\text{ei}\lvert\frac{1}{2\pi}\int d\omega K_\text{jk}(\omega)e^{-i\omega \tau_\text{jk}}\rvert^2.
\end{align}
\end{gather}
\end{widetext}
The delay time $\tau_\text{jk}$ is defined as the difference in emission times between the signal and idler photons: $\tau_\text{jk}\equiv t_\text{sj}-t_\text{ik}$. If we define $\textbf{T}\equiv \textbf{M}^{-1}$, then the kernel functions $K_\text{jk}\left(\omega\right)$ are defined as
\begin{widetext}
\begin{gather}
\begin{align}
K_\text{ff}\left(\omega\right) &\equiv T^*_{31}(\omega) \left[\Gamma_\text{ts} T_{11}(\omega)-1\right] + \Gamma_\text{ts}T_{12}(\omega)T^*_{32}(\omega), \\
K_\text{fb}\left(\omega\right) &\equiv T^*_{41}(\omega) \left[\Gamma_\text{ts} T_{11}(\omega)-1\right] + \Gamma_\text{ts}T_{12}(\omega)T^*_{42}(\omega), \\
K_\text{bf}\left(\omega\right) &\equiv T^*_{32}(\omega) \left[\Gamma_\text{ts} T_{22}(\omega)-1\right] + \Gamma_\text{ts}T_{21}(\omega)T^*_{31}(\omega), \\
K_\text{bb}\left(\omega\right) &\equiv T^*_{42}(\omega) \left[\Gamma_\text{ts} T_{22}(\omega)-1\right] + \Gamma_\text{ts}T_{21}(\omega)T^*_{41}(\omega)
\end{align}
\end{gather}
\end{widetext}
where $T_\text{ln}$ are the elements of the matrix $\textbf{T}$. Equations 30-33 can be evaluated by inverting the matrix $\textbf{M}$ to obtain the elements of $\textbf{T}$. From there, Equation 29 can be evaluated for all j=f,b and k=f,b. The resulting second-order cross-correlation functions are
\begin{widetext}
\begin{gather}
\begin{align}
p_\text{ff}(\tau)&= 
\begin{cases} 
N e^{\Gamma_\text{ti} \tau} \Big\lvert \left[c_0 f e^{-i\theta}-c_1 b \right] \cos \left( \beta_\text{i} \tau \right)+ \left[ c_2 f e^{-i \theta} + c_3 b\right] \sin \left(\beta_\text{i} \tau \right) \Big\rvert^2 & \left(\tau < 0\right), \\
N e^{-\Gamma_\text{ts} \tau} \Big\lvert \left[c_0 f e^{-i\theta}-c_1 b \right] \cos \left( \beta_\text{s} \tau \right)- \left[ c_3 f e^{-i \theta} + c_2 b\right] \sin \left(\beta_\text{s} \tau \right) \Big\rvert^2  & \left(\tau \geq 0 \right),
\end{cases}
\\
p_\text{fb}(\tau)&= 
\begin{cases} 
N e^{\Gamma_\text{ti} \tau} \Big\lvert \left[c_2 f e^{-i\theta}+c_3 b \right] \cos \left( \beta_\text{i} \tau \right)- \left[ c_0 f e^{-i \theta} - c_1 b\right] \sin \left(\beta_\text{i} \tau \right) \Big\rvert^2 & \left(\tau < 0\right), \\
N e^{-\Gamma_\text{ts} \tau} \Big\lvert \left[c_2 f e^{-i\theta}+c_3 b \right] \cos \left( \beta_\text{s} \tau \right)- \left[ c_1 f e^{-i \theta} - c_0 b\right] \sin \left(\beta_\text{s} \tau \right) \Big\rvert^2  & \left(\tau \geq 0 \right),
\end{cases}
\\
p_\text{bf}(\tau)&= 
\begin{cases} 
N e^{\Gamma_\text{ti} \tau} \Big\lvert \left[c_3 f e^{-i\theta}+c_2 b \right] \cos \left( \beta_\text{i} \tau \right)+ \left[ c_1 f e^{-i \theta} - c_0 b\right] \sin \left(\beta_\text{i} \tau \right) \Big\rvert^2 & \left(\tau < 0\right), \\
N e^{-\Gamma_\text{ts} \tau} \Big\lvert \left[c_3 f e^{-i\theta}+c_2 b \right] \cos \left( \beta_\text{s} \tau \right)+ \left[ c_0 f e^{-i \theta} - c_1 b\right] \sin \left(\beta_\text{s} \tau \right) \Big\rvert^2  & \left(\tau \geq 0 \right),
\end{cases}
\\
p_\text{bb}(\tau)&= 
\begin{cases} 
N e^{\Gamma_\text{ti} \tau} \Big\lvert \left[-c_1 f e^{-i\theta}+c_0 b \right] \cos \left( \beta_\text{i} \tau \right)+ \left[ c_3 f e^{-i \theta} + c_2 b\right] \sin \left(\beta_\text{i} \tau \right) \Big\rvert^2 & \left(\tau < 0\right), \\
N e^{-\Gamma_\text{ts} \tau} \Big\lvert \left[-c_1 f e^{-i\theta}+c_0 b \right] \cos \left( \beta_\text{s} \tau \right)- \left[ c_2 f e^{-i \theta} + c_3 b\right] \sin \left(\beta_\text{s} \tau \right) \Big\rvert^2  & \left(\tau \geq 0 \right),
\end{cases}
\end{align}
\end{gather}
\end{widetext}
where $f\equiv\lvert a_\text{pf}\left(\Delta\right)\rvert^2$ and $b\equiv\lvert a_\text{pb}\left(\Delta\right)\rvert^2$ are the detuning-dependent energies of the intracavity forward and backward propagating pump fields, respectively. We define a new phase $\theta\equiv 2\theta_\text{p} +\phi_\beta$ where $\theta_\text{p}$ is the relative phase between forward and backward propagating pump modes ($\theta_\text{pf}-\theta_\text{pb}$) and $\phi_\beta\equiv 2\phi_{\beta_\text{p}}-\phi_{\beta_\text{s}}-\phi_{\beta_\text{i}}$. The other coefficients are given by
\begin{widetext}
\begin{gather}
\begin{align}
N &= \frac{4\Gamma_\text{es}\Gamma_\text{ei}\Gamma_\text{t}^4 g^2}{\left[\left(c_0-c_1\right)\left(c_0+c_1\right)\right]^2}\\
c_0 &= i \left(4\Delta + i\Gamma_{t}\right) \left[-4 \beta_\text{s}^2-4 \beta_\text{i}^2+\left(4\Delta + i\Gamma_\text{t}\right)^2\right] \\
c_1 &= -8i\beta_\text{s}\beta_\text{i}\left(4\Delta + i\Gamma_{t}\right) \\
c_2 &= -2 \beta_\text{i} \left[4 \beta_\text{s}^2-4 \beta_\text{i}^2+\left(4\Delta + i\Gamma_\text{t}\right)^2\right] \\
c_3 &= -2 \beta_\text{s} \left[-4 \beta_\text{s}^2+4 \beta_\text{i}^2+\left(4\Delta + i\Gamma_\text{t}\right)^2\right]
\end{align}
\end{gather}
\end{widetext}
where $\Gamma_\text{t}\equiv\Gamma_\text{ts}+\Gamma_\text{ti}$ and $N$ provides normalization. We note that $c_0,c_1,c_2,$ and $c_3$ are detuning-dependent, giving a strong detuning-dependence to Equations 34-37. We have ignored terms higher than first order in $f$ and $b$ on account of operating in the weak interaction limit. \\

\subsection{B. Theoretical Reversal of the Pump Propagation Direction}
\indent Equations 2-5 are developed from Equations 22-23 of the previous section. We define
\begin{align}
\kappa(\Delta) \equiv \frac{-i\sqrt{\Gamma_\text{ep}}}{\left(i\Delta-\Gamma_\text{tp}/2\right)^2+\lvert\beta_\text{p}\rvert^2}
\end{align}
such that Equations 22-23 can be written more succinctly as
\begin{align}
a_\text{pf}\left(\Delta\right)&=\kappa(\Delta) \left[\left(i\Delta-\Gamma_\text{tp}/2\right)b_\text{pf}-i\beta_\text{p}b_\text{pb}\right],\\
a_\text{pb}\left(\Delta\right)&=\kappa(\Delta) \left[\left(i\Delta-\Gamma_\text{tp}/2\right)b_\text{pb}-i\beta^*_\text{p}b_\text{pf}\right].
\end{align}
The directions forward and backward are here defined relative to the location of pump port 1 and so a change of notation gives
\begin{align}
a_\text{cw}\left(\Delta\right)&=\kappa(\Delta) \left[\left(i\Delta-\Gamma_\text{tp}/2\right)b_1-i\beta_\text{p}b_2\right],\\
a_\text{ccw}\left(\Delta\right)&=\kappa(\Delta) \left[\left(i\Delta-\Gamma_\text{tp}/2\right)b_2-i\beta^*_\text{p}b_1\right]
\end{align}
where cw denotes clockwise propagation within the optical microdisk, ccw denotes counter-clockwise propagation, and 1 and 2 refer to pump fields introduced from port 1 and port 2, respectively. \\
\indent If the device is pumped via port 1 only, then Equations 46-47 become
\begin{align}
a_\text{cw}\left(\Delta\right)&=\kappa(\Delta) \left(i\Delta-\Gamma_\text{tp}/2\right)b_1,\\
a_\text{ccw}\left(\Delta\right)&=\kappa(\Delta) \left(-i \beta^*_\text{p}\right)b_1.
\end{align}
Likewise, if the pump is introduced only at port 2, 
\begin{align}
a_\text{cw}\left(\Delta\right)&=\kappa(\Delta)\left(-i\beta_\text{p}\right)b_2,\\
a_\text{ccw}\left(\Delta\right)&=\kappa(\Delta) \left(i\Delta-\Gamma_\text{tp}/2\right)b_2.
\end{align}
Recalling that the pump propagation direction is always defined relative to port of introduction (forward propagation is movement away from the port and backward propagation is movement toward the port), Equations 48-49 are equivalently
\begin{align}
a_\text{pf}\left(\Delta\right)&=\kappa(\Delta) \left(i\Delta-\Gamma_\text{tp}/2\right)b_1,\\
a_\text{pb}\left(\Delta\right)&=\kappa(\Delta) \left(-i \beta^*_\text{p}\right)b_1.
\end{align}
Equations 50-51 can likewise be written (defined relative to pump port 2) as
\begin{align}
a_\text{pf}\left(\Delta\right)&=\kappa(\Delta) \left(i\Delta-\Gamma_\text{tp}/2\right)b_2, \\
a_\text{pb}\left(\Delta\right)&=\kappa(\Delta)\left(-i\beta_\text{p}\right)b_2.
\end{align}
Recognizing that $\beta_\text{p}\equiv \lvert \beta_\text{p}\rvert e^{i\phi_\text{p}}$, it is clear that Equations 53 and 55 differ by a phase factor $e^{2i\phi_\text{p}}$. Thus pumping at the second port instead of the first alters the relative phase between intracavity modes by $2\phi_p$, twice the pump scattering phase. The phase shift can be observed in the intracavity quantum state \cite{CommPhysLinGroupPaper}
\begin{align}
&\lvert \Psi (t) \rangle_\text{si} = \nonumber \\
 &\Big( c_\text{f} \cos^2(\vert \beta\vert t) -c_\text{b} e^{\pm 2i\phi_\text{p}}\sin^2(\vert \beta \vert t) \Big) \lvert f \rangle_\text{s} \lvert f \rangle_\text{i}+ \nonumber \\
&\Big( c_\text{b} e^{\pm 2i\phi_\text{p}} \cos^2(\vert \beta\vert t) -c_\text{f}\sin^2(\vert \beta \vert t) \Big) \lvert b \rangle_\text{s} \lvert b \rangle_\text{i}+ \nonumber \\
&i\left(c_\text{f}+c_\text{b} e^{\pm 2i\phi_\text{p}}\right)\sin(2\vert\beta\vert t) \Big[ \lvert f \rangle_\text{s} \lvert b \rangle_\text{i} + \lvert b \rangle_\text{s} \lvert f \rangle_\text{i}\Big]
\end{align}
where $c_\text{f}$ and $c_\text{b}$ are proportional to the squares of the respective complex field amplitudes. Dependence on the phase $\phi_p$ has been removed from these coefficients to be shown explicitly. The sign of the phase exponent $2i\phi_p$ is negative when the device is pumped from port 1 and positive when pumped from port 2. The phase difference will influence the quantum interference occurring within the microresonator and induce corresponding phase shifts in the second-order biphoton cross-correlations.\\
\subsection{C. Experimental Setup}
\indent The silicon microdisk used to demonstrate nonreciprocity was fabricated on a silicon on insulator (SOI) wafer with silicon thickness of 260 nm and buried oxide thickness of 2 $\mu$m. The electron beam resist ZEP520A was spun onto the wafer and the pattern was etched into the resist by a JEOL JBX-9500FS electron beam lithography system. The etched pattern was transferred from the resist to the silicon layer by an Oxford Cobra inductively coupled plasma (ICP) reactive ion etcher (RIE) with an $\text{SF}_6$/$\text{C}_4\text{F}_8$ gas chemistry. A wet etch in hydrofluoric acid (HF) was then performed to remove the buried oxide layer.\\
\indent The experimental setup used to observe nonreciprocity in the microdisk is shown in Fig.~5(a-b). A tapered optical fiber is introduced to two forks fabricated near the microdisk, such that light may be evanescently coupled from the tapered fiber to the device. A tunable continuous-wave (CW) laser is split by a 90:10 beamsplitter (BS). The smaller portion of the light passes through a Mach-Zehnder interferometer (MZI) to calibrate measurement of the cavity resonance. The other 90\% of the laser light provides the pump to the device after first passing through a fiber-polarization controller (FPC) and course-wavelength-division multiplexer (CWDM). The FPC allows the polarization to be tuned for optimal coupling to the device. The CWDM provides channels for counter-propagating single-photons while also protecting all single photon channels from pollution by amplified spontaneous emission (ASE). A second identical CWDM is positioned at the opposite port of the tapered optical fiber such that light exiting the device by this port can be separated into signal, idler, and pump channels. The pump channel is directed to a fast photodiode, providing a transmission measurement by which the laser can be locked to the optical cavity. Signal and idler channels from both CWDM are each directed to an optical switch, with both signal channels directed to one switch and both idler channels directed to another. The optical switches give full control over which pair of channels will be measured. The idler and signal photons then each pass to separate narrowband tunable bandpass filters (TBFs) to eliminate Raman scattering noise from single-photon channels. The single photons then pass through FPCs before detection by superconducting nanowire single-photon detectors (SNSPDs). The FPCs are used to optimize the detection efficiencies of the SNSPDs, which are polarization dependent. With polarization optimized, the SNSPDs have detection efficiencies of 53\%. Their low timing jitter of 17 ps allows for highly temporally resolved measurements. The detection of every photon time relative to the start of measurement is recorded (time-tagged) by a time-correlated single-photon counter (TCSPC). The time-binning of the photon counter is 4 ps, and the total data acquisition time is 180 s for the pump doublet data set (Fig.~2-3) and 900 s for the pump singlet data set (Fig.~4). 
\begin{figure*}
\begin{center}
\includegraphics[width=2\columnwidth]{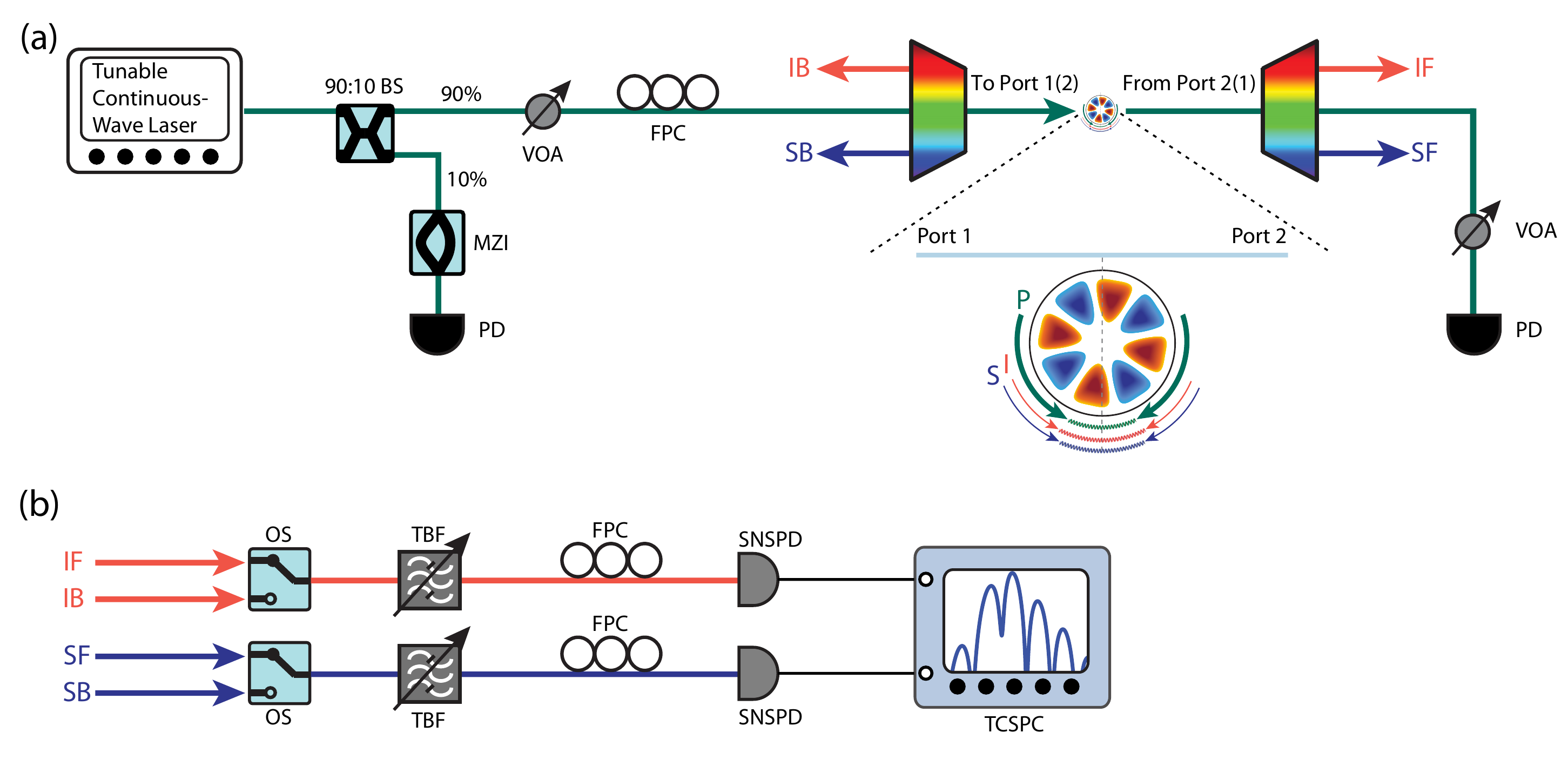}
\caption{Experimental setup. (a) Photon pair generation by spontaneous four-wave mixing (SFWM). A tunable continuous-wave (CW) laser provides the pump. This pump is split by a 90:10 beamsplitter (BS) to pass 10\% of the laser light through a Mach-Zehnder interferometer (MZI) to a photodiode (PD). This light is used as a reference to accurately measure the cavity resonance. The other 90\% of the laser light passes through a variable optical attenuator (VOA) and a fiber polarization controller (FPC). A course-wavelength-division multiplexer (CWDM) eliminates amplified spontaneous emission (ASE) from the pump and provides channels for counter-propagating single photons. The pump is introduced to the silicon microdisk via either port 1 or port 2 of a tapered optical fiber. A second CWDM sits at the output of the opposite port, and again provides single-photon channels. Transmitted pump light passes through a VOA to a PD that is used to lock the laser to the optical cavity. (b) Measurement of the generated photon pairs. Idler photons and signal photons are sent to a pair of optical switches (OS) that allow for measurement of all pairs of single-photon channels. Tunable bandpass filters (TBF) eliminate Raman scattering noise from the channels. FPCs are used to adjust polarization for optimal detection efficiency by superconducting nanowire single-photon detectors (SNSPD). A time-correlated single-photon counter (TCSPC) records the detection time of each photon.}
\end{center}
\end{figure*}
\FloatBarrier

\bibliography{NonrecipV4}

\end{document}